\documentclass[11pt]{article}

\usepackage{graphicx}
\usepackage{amsmath}
\usepackage{amssymb}
\usepackage{latexsym}
\numberwithin{equation}{section}

\usepackage{graphicx}

\begin{document}

\title{Hawking radiation for Dirac spinors on the ${\mathbb{RP}}^3$ geon}
\author{Paul Langlois\footnote{Electronic Address: pmxppl@nottingham.ac.uk} 
\\ \textit{School of Mathematical Sciences, University of Nottingham} 
\\ 
\textit{Nottingham NG7 2RD, UK}}
\date{November 2004} 
\maketitle

\begin{abstract}
We analyse the Hawking(-Unruh) effect for a massive Dirac spinor on
the $\mathbb{Z}_2$ quotient of Kruskal spacetime known as the
${\mathbb{RP}}^3$ geon. There are two distinct Hartle-Hawking-like
vacua, depending on the choice of the spin structure, and suitable
measurements in the static region (which on its own has only one spin
structure) distinguish these two vacua. However, both vacua appear
thermal in the usual Hawking temperature to certain types of
restricted operators, including operators with support in the
asymptotic future (or past). Similar results hold in a family of
topologically analogous flat spacetimes, where we show the two vacua
to be distinguished also by the shear stresses in the zero-mass
limit. As a by-product, we display the explicit Bogolubov transformation
between the Rindler-basis and the Minkowski-basis for massive
Dirac fermions in four-dimensional Minkowski spacetime. 
\end{abstract}

\section{Introduction}

It has been known for nearly thirty years that a black hole formed by
star collapse will radiate thermally at the Hawking
temperature~\cite{h:rad}. This discovery led to black hole
thermodynamics and in particlar to the acceptance of one quarter of
the area \cite{b:bht} as the physical entropy of the hole. It was
realised shortly after the collapsing star analysis that the same
temperature and entropy can also be obtained by considering the
Kruskal-Szekeres extension of Schwarzschild and on it a quantum
state that describes a black hole in thermal equilibrium with its
environment
\cite{hh-vacuum,israel-vacuum,u:unruh,GH1,bd:book,kay-wald,wald-qft}. The
defining characteristics of this Hartle-Hawking state are that it is
regular everywhere on the Kruskal manifold and invariant under the
continuous isometries~\cite{kay-wald,wald-qft}.

From the physical point of view, a puzzling feature of the
Hartle-Hawking state is its reliance on the whole Kruskal
manifold. The manifold has two static regions, causally disconnected
from each other and separated by a bifurcate Killing horizon, but the
thermal properties manifest themselves when the state is probed in
only one of the static regions. To explore the significance
of the second exterior region, Louko and Marolf
\cite{lm:geon} investigated scalar field quantisation on the spacetime
known in the terminology of \cite{fw:topcen} as the ${\mathbb{RP}}^3$
geon (for earlier work on the classical properties of this spacetime,
see \cite{mw:geon,Gi:thesis}). The ${\mathbb{RP}}^3$ geon is a
$\mathbb{Z}_2$ quotient of Kruskal, it is space and time orientable,
it contains a black and white hole, but it only has one static region,
isometric to standard exterior Schwarzschild. It was shown in 
\cite{lm:geon} that the Hartle-Hawking like quantum state on the geon 
does not appear thermal to all observers in the
exterior region, but it does appear thermal in the standard Hawking
temperature when probed by suitably restricted operators. In
particular, the state appears thermal in the standard temperature to
every operator far from the hole and with support at asymptotically
late (or early) Schwarzschild times.

The purpose of this paper is to extend the scalar field analysis of
\cite{lm:geon} to massive Dirac fermions. The main new issue with
fermions is that while exterior Schwarzschild and Kruskal both have
spatial topology ${\mathbb{R}} \times S^2$ and hence a unique spin
structure, the ${\mathbb{RP}}^3$ geon has spatial topology
${\mathbb{RP}}^3 \setminus \{ \textrm{point} \}$ and admits two
inequivalent spin structures. The geon thus has \emph{two\/}
Hartle-Hawking like states for fermions, one for each spin
structure. Our first aim is to examine whether these states appear
thermal when probed in the exterior region: We shall find that they
do, in a limited sense similar to what was found for the scalar field
in~\cite{lm:geon}. Our second aim is to examine whether these two
states can be distinguished by observations limited to the exterior
region. We shall find that they can be in principle distinguished by
suitable interference experiments: The states contain Boulware-type
excitations in correlated pairs, and the spin structure affects the
relative phase between the members of each pair. This means that the
restriction of the Hartle-Hawking type state to the geon exterior not
only tells that the classical geometry behind the horizons differs
from Kruskal but also is sensitive to a quantisation ambiguity whose
existence cannot be deduced from the exterior geometry. In this sense,
probing the quantum state in the exterior region reveals in principle
both classical and quantum information from behind the horizons. How
this information might be uncovered in practice, for example by
particle detectors with a local coupling to the fermion field, 
presents an interesting question for future work. 

We analyse the same issues also on a family of Rindler spaces whose
topology mimics that of the geon~\cite{lm:geon}. While the results are
qualitatively similar to those on the geon, the effects of the spin
structure appear in a much more transparent form, and these Rindler
spaces thus offer a testing ground for localised particle detector
models that aim to resolve the phase factors determined by the spin
structure. We further compute the Rindler-space 
stress-energy tensor explicitly in the 
massless limit, showing that on the geon-type Rindler space the
spatial orientation determined by the spin structure can be detected
from a nonvanishing shear part of the stress-energy. 
As a by-product, we obtain the Bogolubov transformation for massive
Dirac fermions on (ordinary) Rindler space in $(3+1)$ dimensions,
which to the knowledge of the author has not appeared in the
literature.\footnote{Thermality for massive fermions on
$(3+1)$-dimensional Rindler space is demonstrated by other methods in
\cite{tk:takagi,cd:candeld}. The massive $(1+1)$-dimensional case is
considered in~\cite{smg:rin}. The massive $(3+1)$ case is addressed in
\cite{or:oriti} but the Rindler modes constructed therein are not
suitably orthonormal in the Dirac inner product.}

The rest of the paper is as follows. Massive fermions on the
topologically nontrivial Rindler
spaces are analysed in 
section~\ref{sec-unruh}, and the massless limit stress-energy 
is computed in section~\ref{sec-Exp}. Massive fermions on the
${\mathbb{RP}}^3$ geon are analysed in section~\ref{sec-hawking}. 
Section \ref{sec-conclu} gives a brief summary and discussion. 

We work througout in natural units, 
$\hbar=c=G=1$. The metric signature is $(+,-,-,-)$. Complex
conjugation is denoted by ${}^*$ and charge conjugation by~${}^c$.

\section{The Unruh effect on $M_0$ and $M_-$ for massive spinors}
\label{sec-unruh}

In this section we discuss the Unruh effect for the massive Dirac
field on two flat spacetimes whose global properties mimic
respectively those of the Kruskal manifold and the ${\mathbb{RP}}^3$
geon. In section \ref{sec-spacet} we recall the construction of these
spacetimes, denoted by $M_0$ and $M_-$~\cite{lm:geon}, and discuss the
spin structures they admit. The Minkowski-like vacua on $M_0$ and
$M_-$ are constructed in section~\ref{sec-quant}, and the
Rindler-particle content of these vacua is found from the explicit
Bogolubov transformation in sections
\ref{sec-bog} and~\ref{sec-bog2}.

\subsection{The spacetimes $M_0$ and $M_-$}
\label{sec-spacet}

Let $M$ denote $(3+1)$-dimensional Minkowski space and let $(t,x,y,z)$
be a standard set of Minkowski coordinates. The metric reads 
\begin{equation}
ds^2=dt^2-dx^2-dy^2-dz^2
\ . 
\end{equation}
The spacetimes $M_0$ and $M_-$ are defined as quotients of $M$ under
the isometry groups generated respectively by the isometries 
\begin{subequations}
\begin{align} 
J_0 : 
&\ 
(t,x,y,z) \mapsto (t,x,y,z+2a),
\\
J_- : 
&\ 
(t,x,y,z) \mapsto (t,-x,-y,z+a),
\end{align}	
\end{subequations}
where $a>0$ is a prescribed constant. 
$M_0$~and $M_-$ are space and
time orientable flat Lorentzian manifolds, globally hyperbolic with
spatial topology $\mathbb{R}^2 \times S^1$, and $M_0$ is a double
cover of~$M_-$. We may understand $M_0$ and $M_-$ to be coordinatised
by the Minkowski coordinates with the identifications 
\begin{subequations}
\begin{align} 
M_0:  
&  
\ \ 
(t,x,y,z)\sim(t,x,y,z+2a),
\\
M_-:  
&  
\ \ 
(t,x,y,z)\sim(t,-x,-y,z+a).
\end{align}
\end{subequations}
For further discussion, see~\cite{lm:geon}. 

Due to the $S^1$ factor in the spatial topology, $M_0$ and $M_-$ each
admit two inequivalent spin structures (see e.g.~\cite{is:spin}). We
need a practical way to describe these spin structures.

Consider first~$M_0$. We refer to the vierbein 
\begin{eqnarray}  
V_0=\partial_t && V_1=\partial_x 
\nonumber 
\\ 
V_2=\partial_y && V_3=\partial_z  
\label{eqn:minkvier} 
\end{eqnarray}
as the standard vierbein on~$M_0$. In the standard vierbein,
the two spin structures amount to imposing respectively periodic and
antiperiodic boundary conditions on the spinors in the
$z$-direction: Labelling the spin structures by the index $\epsilon
\in \{1,-1\}$, this means 
\begin{equation} 
\psi(t,x,y,z+2na)
=\epsilon^n
\psi(t,x,y,z), 
\end{equation} 
where $n\in{\mathbb{Z}}$ and $\epsilon=+1$ for the periodic spinors
and $\epsilon=-1$ for the antiperiodic spinors. 

An alternative useful vierbein on $M_0$ is
\begin{eqnarray} 
V_0 & = & 
\partial_t 
\nonumber 
\\
V_1 & = & 
\cos({{\pi}z}/a)\partial_x+\sin({{\pi}z}/a)\partial_y 
\nonumber 
\\
V_2 & = & 
-\sin({{\pi}z}/a)\partial_x+\cos({{\pi}z}/a)\partial_y 
\nonumber 
\\
V_3 & = & 
\partial_z  \ \ ,
\label{eqn:rotet}
\end{eqnarray} 
which rotates counterclockwise by $2\pi$ in the $(x,y)$ plane as $z$
increases by~$2a$. Spinors that are periodic in the standard vierbein
(\ref{eqn:minkvier}) are antiperiodic when written in the rotating
vierbein (\ref{eqn:rotet}) and vice versa. One could further introduce
a vierbein that rotates clockwise by $2\pi$ in the $(x,y)$ plane as
$z$ increases by~$2a$ [(replace $\pi$ with $-\pi$ in
(\ref{eqn:rotet})], but periodic (respectively antiperiodic) boundary
conditions in this vierbein are equivalent to periodic (antiperiodic)
boundary conditions in vierbein~(\ref{eqn:rotet}). This shows that neither
spin structure on $M_0$ involves a preferred spatial orientation.


Consider then~$M_-$. The standard vierbein (\ref{eqn:minkvier}) is not
invariant under $J_-$ and does therefore not provide a
globally-defined vierbein on~$M_-$. However, both the
counterclocwise-rotating vierbein (\ref{eqn:rotet}) and its
clockwise-rotating analogue are invariant under $J_-$ and hence well
defined on~$M_-$. We may therefore specify the two spin structures on
$M_-$ by working in the vierbein (\ref{eqn:rotet}) and imposing
respectively periodic and antiperiodic boundary conditions
under~$J_-$, or equivalently working in the clockwise-rotating
vierbein and interchanging the periodic and antiperiodic boundary
conditions. This shows that the choice of a spin structure on $M_-$
determines a preferred spatial orientation. For concreteness, we shall
specify the spin structure with respect to the
vierbein~(\ref{eqn:rotet}).

\subsection{Minkowski-like vacua}
\label{sec-quant}

In this subsection we quantise a free Dirac field $\psi$ with mass
$m\geq{0}$ on $M_0$ and $M_-$, constructing Minkowski-like vacua
defined with respect to the global timelike Killing
vector~$\partial_t$.


In a general curved spacetime the spinor Lagrangian density is given
in the vierbein formalism by \cite{bd:book}
\begin{equation} 
\mathcal{L}
=
{\det V}
\left(
\tfrac{1}{2}i \bigl[ 
\bar{\psi}\gamma^{\mu}\nabla_{\mu}\psi
-(\nabla_{\mu}\bar{\psi})\gamma^{\mu}\psi
\bigr]
-m\bar{\psi}\psi
\right),
\label{eq:lagdens}
\end{equation}
where $V^{\mu}_{\alpha}$ is a vierbein, 
$V_\alpha=V^{\mu}_\alpha\partial_{\mu}$, 
$\gamma^{\alpha}$ are the flat space Dirac matrices, and 
$\gamma^{\mu}=V^{\mu}_{\alpha}\gamma^{\alpha}$ are the curved space
Dirac matrices satisfying 
$ \{\gamma^{\mu},\gamma^{\nu}\}=2g^{\mu\nu} $. 
The spinor covariant derivative is 
$\nabla_{\alpha}=V^{\mu}_{\alpha}(\partial_{\mu}+\Gamma_{\mu})$ 
with
$\Gamma_{\mu}
=
\tfrac{1}{8}V^{\nu}_{\alpha}V_{\beta\nu;\mu}
[\gamma^{\alpha},\gamma^{\beta}] $. 
Variation of the action 
yields the covariant Dirac equation
\begin{equation}
\label{eqn:gendir}
i\gamma^{\mu}\nabla_{\mu}\psi-m\psi=0 \ .
\end{equation}

It will be useful to work in the local Minkowski coordinates
$(t,x,y,z)$ and in the rotating vierbein~(\ref{eqn:rotet}), which is
well-defined on both $M_0$ and~$M_-$. The Dirac equation 
(\ref{eqn:gendir}) then becomes 
\begin{align}
&
i\bigl\{
\gamma^0\partial_t+\gamma^1
\bigl[ \cos(\pi z/a)\partial_x+\sin(\pi z/a)\partial_y \bigr]
+\gamma^2
\bigl[ -\sin(\pi z/a)\partial_x
+\cos(\pi z/a)\partial_y \bigr]
\nonumber
\\
\noalign{\medskip}
& 
\ \ \ 
+\gamma^3
\bigl[ \partial_z 
-
\tfrac14 (\pi/a)
(\gamma^1\gamma^2-\gamma^2\gamma^1)
\bigr] 
+im
\bigr\}
\psi=0 \ . 
\label{eqn:dir}
\end{align}
The inner product is
\begin{equation}\langle{\psi_1,\psi_2}\rangle
=\int_{t=\mathrm{const}}{dx}\,{dy}\,{dz}\,\,\psi_1^\dagger{\psi_2} \ .
\end{equation}
We denote the inner products
on $M_0$ and $M_-$ by $\langle{\psi_1,\psi_2}\rangle_0$ and
$\langle{\psi_1,\psi_2}\rangle_-$ respectively.

Consider first~$M_0$. To construct solutions to (\ref{eqn:dir}) that
are positive and negative frequency with respect to $\partial_t$, we
begin with the Minkowski space positive and negative frequency
solutions in the standard vierbein (\ref{eqn:minkvier}) (see for
example~\cite{bj:plane,it-zub}) and transform to the rotating vierbein
(\ref{eqn:rotet}) by the spinor transfrmation associated with a
rotation by $\pi$ in the $(x,y)$ plane as $z$ increases
by~$a$. Working here and throughout this section in the standard
representation of the $\gamma$ matrices~\cite{it-zub}, $\gamma^0 =
\bigl(
\begin{smallmatrix} 1 & 0 \\ 0 & -1
\end{smallmatrix} 
\bigr)$ 
and 
$\gamma^i = 
\bigl(
\begin{smallmatrix} 
0 & \sigma_i \\ -\sigma_i & 0 
\end{smallmatrix}
\bigr)$, where $\sigma_i$ are the Pauli matrices, this transformation
reads 
\begin{equation} 
\psi\mapsto{e^{-\frac{\gamma^1\gamma^2\pi{z}}{2a}}\psi}
=
\mathrm{diag} 
\left(
e^{\frac{i\pi{z}}{2a}} , 
e^{\frac{-i\pi{z}}{2a}} , 
e^{\frac{i\pi{z}}{2a}} , 
e^{\frac{-i\pi{z}}{2a}}
\right)
\psi \ .
\end{equation}
The periodic and antiperiodic boundary conditions will then restrict
the momentum in the $z$-direction. We find that a complete set of
normalised positive frequency solutions is $\{U_{j,k_x,k_y,k_z}\}$,
where
\begin{equation}
\label{eqn:minkmodes} 
U_{j,k_x,k_y,k_z}=\frac{1}{4\pi}\sqrt{\frac{(\omega+m)}{a\omega}}
e^{-i\omega{t}+ik_{x}x+ik_{y}y+ik_{z}z}
\, u_{j,k_x,k_y,k_z} \ ,
\end{equation}
with
\begin{equation}
u_1=
\left(
\begin{array}{c}
e^{\frac{i\pi{z}}{2a}}
\\ 
0 
\\ 
e^{\frac{i\pi{z}}{2a}}\frac{k_z}{\omega+m} 
\\
e^{-\frac{i\pi{z}}{2a}}\frac{k_+}{\omega+m} 
\end{array} 
\right)
\ \ \ , \ \ \ 
u_2=
\left(
\begin{array}{c}
0 
\\ 
e^{-\frac{i\pi{z}}{2a}} 
\\ 
e^{\frac{i\pi{z}}{2a}}\frac{k_-}{\omega+m}
\\ 
e^{-\frac{i\pi{z}}{2a}}\frac{-k_z}{\omega+m} 
\end{array} 
\right) \ \ , 
\end{equation}
$k_{\pm}=k_x\pm{i}k_y$ and 
$\omega=\sqrt{m^2+k_x^2+k_y^2+k_z^2}$. 
For spinors that are periodic in the standard
vierbein~(\ref{eqn:minkvier}), 
$k_z = n\pi/a$ with $n\in{\mathbb{Z}}$,
and in the other spin structure 
$k_z= ( n+\tfrac12 )\pi/a$ with $n\in{\mathbb{Z}}$. 
$k_x$~and $k_y$ take all real values. 
The orthonormality relation is 
\begin{equation}
\langle{U_{i,k_x,k_y,k_z},U_{j,k'_x,k'_y,k'_z}}\rangle_0
=\delta_{ij}\delta_{nn'}\delta(k_x-k'_x)\delta(k_y-k'_y) \ .
\end{equation}
Note that the corresponding delta-normalised modes on Minkowski are
$\sqrt{\frac{a}{\pi}}U_{j,k_x,k_y,k_z}$ with $k_z \in \mathbb{R}$. 

Consider then~$M_-$. As $M_0$ is a double cover of~$M_-$, a complete
set of positive frequency modes is obtained by superposing the
modes (\ref{eqn:minkmodes}) and their images under $J_-$ with
phase factors that lead to the appropriate (anti-)periodicity
properties. We choose the set $\{V_{j,k_x,k_y,k_z}\}$ given by 
\begin{eqnarray} 
&&V_{1,k_x,k_y,k_z}=U_{1,k_x,k_y,k_z}
+\epsilon{i}{e^{ik_za}}U_{1,-k_x,-k_y,k_z} \ ,
\nonumber 
\\
&&V_{2,k_x,k_y,k_z}=U_{2,k_x,k_y,k_z}
-\epsilon{i}{e^{ik_za}}U_{2,-k_x,-k_y,k_z} \ ,
\end{eqnarray}
where $k_z= ( n+\tfrac12 )\pi/a$, $n\in{\mathbb{Z}}$ and $\epsilon=1$
($\epsilon=-1$) gives spinors that are periodic (antiperiodic) in the
rotating vierbein~(\ref{eqn:rotet}). As with the scalar field
\cite{lm:geon} there is a redundency in these $V$-modes in that
$V_{j,k_x,k_y,k_z}$ and $V_{j,-k_x,-k_y,k_z}$ are proportional, and we
understand this redundancy to be eliminated by taking for example
$k_y>0$. The orthonormality condition reads
\begin{equation}
\label{eqn:orthV}
\langle{V_{i,k_x,k_y,k_z},V_{j,k'_x,k'_y,k'_z}}\rangle_-
=
\delta_{ij}\delta_{nn'}\delta(k_x-k'_x)\delta(k_y-k'_y) \ .
\end{equation}
Note that $k_z$ takes in both spin structures the same set of values,
which coincides with the set in the twisted spin structure
on~$M_0$.

Given these mode sets, we can canonically quantise in the usual way,
expanding the field in the modes and imposing the usual
anticommutation relations on the coefficients. Let $|0\rangle$ be the
usual Minkowski vacuum on~$M$, defined by the set
$\{U_{j,k_x,k_y,k_z}\}$. We denote by $|0_0\rangle$ the vacuum on
$M_0$ defined by the set $\{U_{j,k_x,k_y,k_z}\}$ with the suitably
restricted values of $k_z$ and by $|0_-\rangle$ the vacuum on $M_-$
defined by the set $\{V_{j,k_x,k_y,k_z}\}$. $|0_0\rangle$~and
$|0_-\rangle$ both depend on the respective spin structures, but in
what follows we will not need an explicit index to indicate this
dependence.

\subsection{Bogolubov transformation on $M_0$}
\label{sec-bog}

In this subsection we find the Rindler-particle content of the
Minkowski-like vacuum on $M_0$ from the explicit Bogolubov
transformation. At the end of the subsection we indicate how the
corresponding results for Minkowski space can be read off from our
formulas.

Let $R_0$ be the right-hand-side Rindler wedge of~$M_0$, $x>|t|$. We
introduce in $R_0$ the usual Rindler coordinates $(\eta,\rho,y,z)$ by 
\begin{align} 
t&= 
\rho\sinh\eta \ ,
\nonumber 
\\ 
x&=
\rho\cosh\eta \ ,
\label{eqn:rin} 
\end{align}  
with $\rho>0$ and $-\infty<\eta<\infty$, understood with the
identification
$(\eta,\rho,y,z)\sim(\eta,\rho,y,z+2a)$. The metric reads 
\begin{equation}
{ds}^2={\rho}^2{d\eta}^2-{d\rho}^2-{dy}^2-{dz}^2
 \ \ . 
\end{equation} 
$R_0$ is a globally hyperbolic spacetime with the complete timelike
Killing vector $\partial_{\eta}=t\partial_x+x\partial_t$, which
generates boosts in the $(t,x)$ plane. The worldines at constant
$\rho$, $y$ and $z$ are those of observers accelerated uniformly in
the $x$-direction with acceleration $\rho^{-1}$ and proper time
$\rho\eta$.

We need in $R_0$ a set of orthonormal field modes that are positive
frequency with respect to~$\partial_\eta$. In the vierbein aligned
along the Rindler coordinate axes,
\begin{equation}
\label{eqn:rinvier}
V^{\mu}_{a}=\mathrm{diag}({\rho}^{-1},1,1,1)
 \ , 
\end{equation}
the Dirac 
equation (\ref{eqn:gendir}) becomes 
\begin{equation}
\label{eqn:dirrin}
(i\partial_\eta
+i\rho\gamma^0\gamma^1\partial_\rho
+i\rho\gamma^0\gamma^2\partial_y
+i\rho\gamma^0\gamma^3\partial_z
+{i\gamma^0\gamma^1}/2-m\rho\gamma^0)
\psi
=0
\ ,
\end{equation}
where the $\gamma$ matrices are the usual flat space
$\gamma{'s}$. We separate (\ref{eqn:dirrin}) 
by an ansatz of
simultaneous eigenfunctions of $-i\partial_y$, $-i\partial_z$ 
and the Rindler Hamiltonian. In 
view of comparison with $M_-$ in subsection~\ref{sec-bog2}, we wish 
the solutions to have simple transformation properties 
under~$J_-$. Modes that achieve this are 
\begin{equation}
\label{eqn:rinsol}
\psi^R_{j,M,k_y,k_z}(\eta,\rho,y,z)
=N_j\left(X^R_jK_{iM-\frac{1}{2}}(k\rho)
+Y^R_jK_{iM+\frac{1}{2}}(k\rho)\right)e^{-iM\eta+ik_yy+ik_zz} \ ,
\end{equation}
where
\begin{align} 
X^R_1
= & \left(\begin{array}{ccccccccccccccccccccc}\frac{k_z}{|k_z|}(k_y-im) 
\\ -i(|k_z|-\kappa) 
\\ -i(|k_z|-\kappa) 
\\ \frac{k_z}{|k_z|}(k_y-im) 
\end{array} 
\right) \ ,
 &           
Y^R_1
= & \left(\begin{array}{ccccccccccccccccccccc}\frac{k_z}{|k_z|}(|k_z|-\kappa) 
\\ i(k_y-im) 
\\ -i(k_y-im)
\\ -\frac{k_z}{|k_z|}(|k_z|-\kappa) 
\end{array} 
\right) \ ,
\nonumber 
\\ 
X^R_2
= & \left(\begin{array}{ccccccccccccccccccccc}\frac{k_z}{|k_z|}(|k_z|-\kappa) 
\\ i(k_y+im) 
\\ i(k_y+im)
\\ \frac{k_z}{|k_z|}(|k_z|-\kappa) 
\end{array} 
\right) \ , 
 &    
Y^R_2
= & \left(\begin{array}{ccccccccccccccccccccc}\frac{k_z}{|k_z|}(k_y+im) 
\\ -i(|k_z|-\kappa) 
\\ i(|k_z|-\kappa) 
\\ -\frac{k_z}{|k_z|}(k_y+im) 
\end{array} 
\right) \ , 
\end{align} 
and 
\begin{align} 
N_1=
& \frac{e^{-\frac{i\pi}{4}}\sqrt{\cosh(\pi{M})(\kappa^2-k_z^2)}}
{4\pi(k_y-im)\sqrt{a\pi(\kappa-|k_z|)}} \ , 
\nonumber
\\ 
\label{eqn:norm123} 
N_2=
& \frac{e^{-\frac{i\pi}{4}}\sqrt{\cosh(\pi{M})(\kappa^2-k_z^2)}}
{4\pi(k_y+im)\sqrt{a\pi(\kappa-|k_z|)}} \ , 
\end{align}
$j=1,2$, $\kappa=(m^2+{k_y}^2+{k_z}^2)^{1/2}$, $M>0$ and
$k_y{\in}{\mathbb{R}}$. In the spin structure where the spinors are
periodic (respectively antiperiodic) in the nonrotating
vierbein~(\ref{eqn:rinvier}), the values of $k_z$ are $n\pi/a$
(respectively $(n+\tfrac12)\pi/a$) with
$n\in{\mathbb{Z}}$. $K_{iM+\frac{1}{2}}$ is a modified Bessel
function~\cite{zw:zwillinger}. For $k_z=0$, we understand
the formulas in (\ref{eqn:rinsol})--(\ref{eqn:norm123}) and in what
follows to stand for their limiting values
as $k_z \to {0_+}$. 
The modes are orthonormal as 
\begin{equation}
\langle{\psi^R_{i,M,k_y,k_z},\psi^R_{j,M',k_y',k_z'}}\rangle_{R_0}
=\delta_{ij}\delta_{nn'}\delta(M-M')\delta(k_y-k_y') 
\ , 
\label{eq:psi-ON}
\end{equation}
where the inner product is (see e.g (\cite{cd:candeld}))
\begin{equation}
\label{eqn:inner}
\langle{\psi_1,\psi_2}\rangle_{R_0} 
=\int{d}\rho\,{dy}\,{dz}\,\psi_1^\dagger{\psi_2} \ ,
\end{equation}
taken on an $\eta=\mathrm{constant}$ hypersurface. 

While the above modes would be sufficient for quantising in $R_0$ in
its own right, they are not suitable for analytic continuation
arguments across the horizons, as the vierbein (\ref{eqn:rinvier})
becomes singular in the limit $x\to |t|$. We therefore express the
modes in the vierbein~(\ref{eqn:rotet}), which is globally defined
on~$M_0$. This vierbein will further make the comparison to $M_-$
transparent in subsection~\ref{sec-bog2}. The Lorentz transformation
between (\ref{eqn:rinvier}) and (\ref{eqn:rotet}) is 
a boost by rapidity $-\eta$ in the
$(\eta,\rho)$ plane followed by a rotation by $\pi$ as $z\mapsto{z+a}$
in the $(x,y)$ plane. The corresponding transformation on the spinors
is
\begin{equation}
\psi
\mapsto 
e^{-\frac{\gamma^1\gamma^2\pi{z}}{2a}}
e^{\frac{\gamma^0\gamma^1\eta}{2}}
\psi  \ . 
\end{equation}
In the vierbein~(\ref{eqn:rotet}), our solutions thus become 
\begin{align}
\psi^R_{j,M,k_y,k_z}(t,x,y,z) 
& =
N_j
\left(
X'^R_jK_{iM-\frac{1}{2}}(k\rho)e^{-(iM-\frac{1}{2})\eta}\right. 
\nonumber
\\
\label{eqn:rinmodes} 
&
\hspace{7ex} 
\left.\mbox{}+Y'^R_jK_{iM+\frac{1}{2}}(k\rho)e^{-(iM+\frac{1}{2})\eta}
\right)
e^{ik_yy+ik_zz}
\ , 
\end{align}
where
\begin{align} 
X'^R_1
=& 
\left(
\begin{array}{ccccccccccccccccccccc}
e^{\frac{i\pi{z}}{2a}}\frac{k_z}{|k_z|}(k_y-im) 
\\ 
-e^{-\frac{i\pi{z}}{2a}}i(|k_z|-\kappa) 
\\ 
-e^{\frac{i\pi{z}}{2a}}i(|k_z|-\kappa) 
\\ 
e^{-\frac{i\pi{z}}{2a}}\frac{k_z}{|k_z|}(k_y-im) 
\end{array} 
\right) \ , 
& 
Y'^R_1=&
\left(
\begin{array}{ccccccccccccccccccccc}
e^{\frac{i\pi{z}}{2a}}\frac{k_z}{|k_z|}(|k_z|-\kappa) 
\\ 
e^{-\frac{i\pi{z}}{2a}}i(k_y-im) 
\\ 
-e^{\frac{i\pi{z}}{2a}}i(k_y-im)
\\ 
-e^{-\frac{i\pi{z}}{2a}}\frac{k_z}{|k_z|}(|k_z|-\kappa) 
\end{array} 
\right) \ , 
\nonumber 
\\
X'^R_2=&
\left(
\begin{array}{ccccccccccccccccccccc}
e^{\frac{i\pi{z}}{2a}}\frac{k_z}{|k_z|}(|k_z|-\kappa) 
\\ 
e^{-\frac{i\pi{z}}{2a}}i(k_y+im) 
\\ 
e^{\frac{i\pi{z}}{2a}}i(k_y+im)
\\ 
e^{-\frac{i\pi{z}}{2a}}\frac{k_z}{|k_z|}(|k_z|-\kappa) 
\end{array} 
\right) \ , 
& 
Y'^R_2=&
\left(
\begin{array}{ccccccccccccccccccccc}
e^{\frac{i\pi{z}}{2a}}\frac{k_z}{|k_z|}(k_y+im) 
\\ 
-e^{-\frac{i\pi{z}}{2a}}i(|k_z|-\kappa) 
\\ 
e^{\frac{i\pi{z}}{2a}}i(|k_z|-\kappa) 
\\ 
-e^{-\frac{i\pi{z}}{2a}}\frac{k_z}{|k_z|}(k_y+im) 
\end{array} 
\right) \ . 
\end{align}

We proceed similarly in the left-hand-side Rindler wedge~$L_0$, $x<
-|t|$. We define the Rindler coordinates in $L_0$ by 
\begin{align}
t&=  -\rho\sinh\eta \ ,
\nonumber
\\ 
x&=  -\rho\cosh\eta \ , 
\label{eqn:rinl} 
\end{align}
again with $\rho>0$ and $-\infty<\eta<\infty$. Note that
$\partial_\eta$ is now past-pointing. In the
vierbein~(\ref{eqn:rotet}), a complete orthonormal set of positive
frequency modes 
with respect to~$\partial_\eta$ is
\begin{align}
\psi^L_{j,M,k_y,k_z}(t,x,y,z) 
& = 
N_j
\left(
X'^L_jK_{iM-\frac{1}{2}}(k\rho)e^{-(iM-\frac{1}{2})\eta}
\right. 
\nonumber
\\
&
\hspace{7ex} 
\left.
\mbox{}+Y'^L_jK_{iM+\frac{1}{2}}(k\rho)e^{-(iM+\frac{1}{2})\eta}
\right)
e^{ik_yy+ik_zz} \ ,
\label{eqn:rinmodesl}
\end{align}
where
\begin{align}
X'^L_1=&
\left(
\begin{array}{ccccccccccccccccccccc}
e^{\frac{i\pi{z}}{2a}}\frac{k_z}{|k_z|}(k_y-im) 
\\ 
-e^{-\frac{i\pi{z}}{2a}}i(|k_z|-\kappa) 
\\ 
-e^{\frac{i\pi{z}}{2a}}i(|k_z|-\kappa) 
\\ 
e^{-\frac{i\pi{z}}{2a}}\frac{k_z}{|k_z|}(k_y-im) 
\end{array} 
\right) \ , 
& 
Y'^L_1=&
\left(
\begin{array}{ccccccccccccccccccccc}
-e^{\frac{i\pi{z}}{2a}}\frac{k_z}{|k_z|}(|k_z|-\kappa) 
\\ 
-e^{-\frac{i\pi{z}}{2a}}i(k_y-im) 
\\ 
e^{\frac{i\pi{z}}{2a}}i(k_y-im)
\\ 
e^{-\frac{i\pi{z}}{2a}}\frac{k_z}{|k_z|}(|k_z|-\kappa) 
\end{array} 
\right) \ , 
\nonumber 
\\
X'^L_2=&
\left(
\begin{array}{ccccccccccccccccccccc}
-e^{\frac{i\pi{z}}{2a}}\frac{k_z}{|k_z|}(|k_z|-\kappa) 
\\ 
-e^{-\frac{i\pi{z}}{2a}}i(k_y+im) 
\\ 
-e^{\frac{i\pi{z}}{2a}}i(k_y+im)
\\ 
-e^{-\frac{i\pi{z}}{2a}}\frac{k_z}{|k_z|}(|k_z|-\kappa) 
\end{array} 
\right) \ , 
&
Y'^L_2=&
\left(
\begin{array}{ccccccccccccccccccccc}
e^{\frac{i\pi{z}}{2a}}\frac{k_z}{|k_z|}(k_y+im) 
\\ 
-e^{-\frac{i\pi{z}}{2a}}i(|k_z|-\kappa) 
\\ 
e^{\frac{i\pi{z}}{2a}}i(|k_z|-\kappa) 
\\ 
-e^{-\frac{i\pi{z}}{2a}}\frac{k_z}{|k_z|}(k_y+im) 
\end{array} 
\right) \ ,
\end{align}
and the 
ranges of $M$, $k_y$ 
and $k_z$ are as in~$R_0$. The orthonormality relation is similar
to~(\ref{eqn:inner}). 

We may now quantise the field in $R_0$ and $L_0$ in the usual
manner. A~complete set of positive frequency modes with respect to the
future-pointing timelike Killing vector is
$\{\psi^R_{j,M,k_y,k_z}(t,x,y,z)\}$ in $R_0$ and 
$\{\psi^L_{j,-M,k_y,k_z}(t,x,y,z)\}$ in $L_0$, both with $M>0$: The
minus sign in $\psi^L$ arises because 
$\partial_\eta$ is past-pointing in~$L_0$. 
The expansion of the field in these modes and their charge conjugates
reads 
\begin{gather}
\Psi=
\sum_{j}\sum_{n}\int_{0}^{\infty}dM\int_{-\infty}^{\infty} 
\!\!\!\!
dk_y 
\left(
a^R_{j,M,k_y,k_z}\psi^{R}_{j,M,k_y,k_z}
+a^L_{j,M,k_y,k_z}\psi^{L}_{j,-M,k_y,k_z} 
\right.
\nonumber
\\     					  
\label{eqn:rinexp}		        
\left.
\mbox{}+b^{R\dagger}_{j,M,k_y,k_z}\psi^{R,c}_{j,M,k_y,k_z}
+b^{L\dagger}_{j,M,k_y,k_z}\psi^{L,c}_{j,-M,k_y,k_z}
\right) \ . 
\end{gather}
As the charge conjugation in our standard representation of the
$\gamma$'s reads $\psi^c =i\gamma^2\psi^*$, where the superscript
${}^*$ stands for normal complex conjugation, it follows that
\begin{align}
\psi^{R,c}_{1,M,k_y,k_z}(t,x,y,z)
= &\ -i\psi^{R}_{1,-M,-k_y,-k_z}(t,x,y,z) \ , 
\nonumber
\\
\label{eq:chargeconjR}
\psi^{R,c}_{2,M,k_y,k_z}(t,x,y,z)
= &\ i\psi^{R}_{2,-M,-k_y,-k_z}(t,x,y,z) \ ,  
\end{align}
with similar expressions for $\psi^L_1$ and $\psi^L_2$. The 
annihilation and creation operators in $R_0$ satisfy the 
usual anticommutation
relations 
\begin{align}
\{a^R_{i,M,k_y,k_z},a^{R\dagger}_{j,M',{k'_y},{k'_z}}\}
= &\ \delta_{ij}\delta_{nn'}\delta(M-M')\delta(k_y-{k'_y}) \ , 
\nonumber
\\
\{b^R_{i,M,k_y,k_z},b^{R\dagger}_{j,M',{k'_y},{k'_z}}\}
= &\ \delta_{ij}\delta_{nn'}\delta(M-M')\delta(k_y-{k'_y}) \ ,
\end{align}
with similar relations holding in $L_0$ and all mixed anticommutators
vanishing. The right and left Rindler vacua, $|0_{R_0}\rangle$ and
$|0_{L_0}\rangle$, are defined as the states annihilated by all the
appropriate annihilation operators. 

Now, we wish to find the Rindler mode content of the Minkowski-like
vacuum $|0_0\rangle$ by Unruh's analytic continuation
method~\cite{u:unruh}. To begin, we continue the sets
$\{\psi^R_{j,M,k_y,k_z}\}$ and $\{\psi^L_{j,-M,k_y,k_z}\}$
analytically to all of~$M_0$, crossing the horizons in the lower
half-plane in complexified~$t$. Using the relations
(\ref{eq:chargeconjR}), their counterparts in $L_0$ and the complex
analytic properties of the Bessel functions~\cite{zw:zwillinger}, we
find that the resulting modes are
\begin{align}
W^{(1)}_{j,M,k_y,k_z}(t,x,y,z)= 
&\ 
\frac{1}{\sqrt{2\cosh(\pi{M})}}
\left(
e^{\frac{\pi{M}}{2}}\psi^R_{j,M,k_y,k_z}
+{e^{-\frac{\pi{M}}{2}}\psi^{L,c}_{j,-M,-k_y,-k_z}}
\right) \ ,
\nonumber
\\ 
W^{(2)}_{j,M,k_y,k_z}(t,x,y,z)= 
&\ 
\frac{1}{\sqrt{2\cosh(\pi{M})}}
\left(
-e^{-\frac{\pi{M}}{2}}\psi^{R,c}_{j,M,-k_y,-k_z}
+{e^{\frac{\pi{M}}{2}}\psi^{L}_{j,-M,k_y,k_z}}
\right) \ ,
\end{align}
where $M>0$. The normalisation is 
\begin{equation}
\langle{W_{i,M,k_y,k_z},W_{j,M',k'_y,k'_z}}\rangle_0
=\delta_{ij}\delta_{nn'}\delta(M-M')\delta(k_y-k'_y) \ .
\end{equation}
We then expand the field as 
\begin{gather}
\Psi=\sum_{j}\sum_{n}
\int_{0}^{\infty}dM
\int_{-\infty}^{\infty} \!\!\!\!dk_y 
\left(
c^{(1)}_{j,M,k_y,k_z}W^{(1)}_{j,M,k_y,k_z}
+c^{(2)}_{j,M,k_y,k_z}W^{(2)}_{j,M,k_y,k_z}
\right. 
\nonumber
\\
\label{eqn:Wexp}
\left.
\mbox{}+d^{(1)\dagger}_{j,M,k_y,k_z}W^{(1),c}_{j,M,k_y,k_z}
+d^{(2)\dagger}_{j,M,k_y,k_z}W^{(2),c}_{j,M,k_y,k_z}
\right) 
\end{gather}
and impose the usual anticommutation relations for the creation and
annihilation operators in~(\ref{eqn:Wexp}). Equating (\ref{eqn:Wexp}) and
(\ref{eqn:rinexp}) and taking inner products with the $\psi$ modes
gives the Bogolubov transformation 
\begin{align}
a^{R}_{j,M,k_y,k_z}= 
&\ 
\frac{1}{\sqrt{2\cosh(\pi{M})}}
\left(
e^{\frac{\pi{M}}{2}}c^{(1)}_{j,M,k_y,k_z}
-{e^{-\frac{\pi{M}}{2}}}d^{(2)\dagger}_{j,M,-k_y,-k_z}
\right) \ ,
\nonumber
\\ 
a^{L}_{j,M,k_y,k_z}= 
&\ 
\frac{1}{\sqrt{2\cosh(\pi{M})}}
\left(
e^{\frac{\pi{M}}{2}}c^{(2)}_{j,M,k_y,k_z}
+{e^{-\frac{\pi{M}}{2}}}d^{(1)\dagger}_{j,M,-k_y,-k_z}
\right) \ ,
\nonumber
\\ 
b^{R,\dagger}_{j,M,k_y,k_z}= 
&\ 
\frac{1}{\sqrt{2\cosh(\pi{M})}}
\left(
e^{\frac{\pi{M}}{2}}d^{(1)\dagger}_{j,M,k_y,k_z}
-{e^{-\frac{\pi{M}}{2}}}c^{(2)}_{j,M,-k_y,-k_z}
\right) \ , 
\nonumber
\\
b^{L,\dagger}_{j,M,k_y,k_z}= 
&\ 
\frac{1}{\sqrt{2\cosh(\pi{M})}}
\left(
e^{\frac{\pi{M}}{2}}d^{(2)\dagger}_{j,M,k_y,k_z}
+{e^{-\frac{\pi{M}}{2}}}c^{(1)}_{j,M,-k_y,-k_z}
\right) \ .
\label{eqn:bog}
\end{align}
As the $W$-modes are by construction purely positive frequency with
respect to $\partial_t$~\cite{u:unruh}, the vacuum annihilated by all
the annihilation operators in (\ref{eqn:Wexp}) is $|0_0\rangle$. This
observation and the transformation (\ref{eqn:bog}) fix the
coefficients in the expansion of $|0_0\rangle$ in terms of the
Rindler-excitations on $|0_{R_0}\rangle$ and~$|0_{L_0}\rangle$. The
final result is
\begin{equation}
\label{eqn:sum}
|0_0\rangle
=\prod_{j,M,k_y,k_z}
\frac{1}{({e^{-2\pi{M}}+1)^{\frac{1}{2}}}}
\sum_{q=0,1}
(-1)^qe^{-\pi{M}q}
|q\rangle^{R}_{j,M,k_y,k_z}
|q\rangle^{L}_{j,M,-k_y,-k_z} 
\ , 
\end{equation}
where the notation on the right hand side is adapted to the tensor
product stucture of the Hilbert space considered: 
\begin{align}
|q\rangle^{R}_{j,M,k_y,k_z}
= &\ (a^{R\dagger}_{j,M,k_y,k_z})^q|0_{R_0}\rangle \ ,
\nonumber
\\
|q\rangle^{L}_{j,M,k_y,k_z}
= &\ (b^{L\dagger}_{j,M,k_y,k_z})^q|0_{L_0}\rangle \ . 
\end{align}

The result (\ref{eqn:sum}) is the massive fermion version of the
familiar bosonic result~\cite{bd:book}, indicating an entangled state
in which the right and left Rindler excitations appear in correlated
pairs.  An operator $\hat{A}^{(1)}$ whose support is in $R_0$ does
not couple to the Rindler-modes in $L_0$ and has hence the expectation
value $\langle0_0|\hat{A}^{(1)}|0_0\rangle=
\mathrm{Tr}(\hat{A}^{(1)}\rho^{(1)})$, where $\rho^{(1)}$ is a
fermionic thermal density matrix in~$R_0$,
\begin{equation}
\rho^{(1)}
=\prod_{j,M,k_y,k_z}
\sum_{q=0,1}
\frac{e^{-2q\pi{M}}}{\sum_{m=0,1}e^{-2m\pi{M}}}
|q\rangle^{R}_{j,M,k_y,k_z}{}^{R}
\langle{q}|_{j,M,k_y,k_z} \ . 
\end{equation}
In particular, the number
operator expectation value takes the fermionic thermal
form, 
\begin{equation}
\label{eqn:number}
\langle0_0|
a^{R\dagger}_{i,M,k_y,k_z}a^{R}_{j,{M'},{k_y}',{k_z}'}
|0_0\rangle
=\frac{1}{(e^{2\pi{M}}+1)}
\delta_{ij}\delta_{n{n'}}\delta(k_y-{k'_y})\delta(M-{M'}) 
\ . 
\end{equation}
The delta-function divergences in (\ref{eqn:number}) arise from the
continuum normalisation of our modes and can be remedied by wave
packets as in the scalar case~\cite{lm:geon}.  This shows that the
Rindler-observers in $R_0$ see $|0_0\rangle$ as a thermal bath at the
usual Unruh temperature, $T= {(2\pi\rho)}^{-1}$. Similar
considerations clearly hold for~$L_0$.

%

The result (\ref{eqn:sum}) incorporates the two spin structures on
$M_0$, $R_0$ and $L_0$, in the allowed values of~$k_z$, and these
values are the same in all our mode sets. The twisted (respectively
untwisted) $|0_0\rangle$ thus induces twisted (untwisted) thermal
states in both $R_0$ and~$L_0$.

To end this subsection, we note that the Bogolubov transformation on
Minkowski space can be read off from our formulas on $M_0$ with minor
systematic changes. There is now only one spin structure and $k_z$
takes all real values. The expressions for the various mode functions
include the additional factor $\sqrt{a/\pi}$, sums over $n$ become
integrals over~$k_z$, and in the normalisation and anticommutation
relations the discrete delta $\delta_{n n'}$ is replaced by the
delta-function $\delta(k_z - k_z')$. The formulas involve still $a$
because the spinors are expressed in the rotating
vierbein~(\ref{eqn:rotet}). Translation into the standard vierbein
(\ref{eqn:minkvier}) can be accomplished by the appropriate spinor
transformation.

\subsection{Bogolubov transformation on $M_-$} 
\label{sec-bog2}

In this subsection we find the Rindler-particle content of the
Minkowski-like vacuum on~$M_-$. 


Let $R_-$ denote the Rindler wedge on~$M_-$, given in our local
coordinates by $x>|t|$. As $R_-$ is isometric to~$R_0$, we may
introduce in $R_-$ the Rindler-coordinates $(\eta,\rho,y,z)$
by~(\ref{eqn:rin}), again with the identifications
$(\eta,\rho,y,z)\sim(\eta,\rho,y,z+2a)$, and quantise as in~$R_0$,
defining the positive and negative frequencies with respect to the
Killing vector~$\partial_\eta$. For convenience of phase factors in
the $W$-modes (\ref{eqn:m-wmode12}) below, we use the mode set
$\{\Psi^R_{j,M,k_y,k_z}\}$, defined as in (\ref{eqn:rinmodes}) except
in that the normalisation factors (\ref{eqn:norm123}) are replaced by
\begin{align}
N_1
= 
&\ 
\frac{e^{-\frac{ik_za}{2}}\sqrt{\cosh(\pi{M})(\kappa^2-k_z^2)}}
{4\pi(k_y-im)\sqrt{a\pi(\kappa-|k_z|)}} \ ,
\nonumber
\\
N_2= 
&\ 
\frac{e^{\frac{ik_za}{2}}\sqrt{\cosh(\pi{M})(\kappa^2-k_z^2)}}
{4\pi(k_y+im)\sqrt{a\pi(\kappa-|k_z|)}} \ .
\end{align}
A key difference from $R_0$ arises, however, from the requirement that
the spinors on $R_-$ must be extendible into spinors in one of the two
spin structures on~$M_-$. By the discussion in
subsection~\ref{sec-spacet}, this implies that $k_z$ is restricted to
the values $k_z=(n+\tfrac12)\pi/a$ with $n\in{\mathbb{Z}}$. Both of
the spin structures on $M_-$ thus induce on $R_-$ the same spin
structure, in which the spinors are antiperiodic in the nonrotating
vierbein~(\ref{eqn:minkvier}).

The inner product on $R_-$ is as in (\ref{eqn:inner}) and 
the orthonormality relation is similar to~(\ref{eq:psi-ON}). 
The mode expansion of the field reads 
\begin{gather}
\Psi
=\sum_{n}\int_{0}^{\infty}dM\int_{-\infty}^{\infty}dk_y
\left(
a_{1,M,k_y,k_z}\Psi^{R}_{1,M,k_y,k_z}
+a_{2,M,k_y,k_z}\Psi^{R}_{2,M,k_y,k_z}
\right.
\nonumber 
\\
\label{eqn:expan-}
\left.
\mbox{}+b^\dagger_{1,M,k_y,k_z}\Psi^{R,c}_{1,M,k_y,k_z}
+b^\dagger_{2,M,k_y,k_z}\Psi^{R,c}_{2,M,k_y,k_z}
\right) \ ,
\end{gather}
where the annihilation and creation operators satisfy the usual
anticommutation relations. The Rindler-vacuum $|0_{R_-}\rangle$ on
$R_-$ is the state annihilated by all the annihilation operators
in~(\ref{eqn:expan-}).

To find the Rindler-mode content of~$|0_-\rangle$, we again use the
analytic continuation method. Working in the local coordinates
$(t,x,y,z)$, we continue the modes $\{\Psi^R_{j,M,k_y,k_z}\}$ across
the horizons in the lower half-plane in complexified $t$ and form the
linear combinations that are globally well-defined on~$M_-$. The
complete set of the resulting $W$-modes is $\{W^R_{j,M,k_y,k_z}\}$,
given by
\begin{align}
&{W^{R}_{1,M,k_y,k_z}(t,x,y,z)}=\frac{1}{\sqrt{2\cosh{\pi{M}}}}
\left(
{e^{\frac{\pi{M}}{2}}}\Psi^{R}_{1,M,k_y,k_z}
+\epsilon{e^{-\frac{\pi{M}}{2}}}\Psi^{R,c}_{2,M,k_y,-k_z}
\right) \ ,
\nonumber
\\
\label{eqn:m-wmode12}
&{W^{R}_{2,M,k_y,k_z}(t,x,y,z)}=\frac{1}{\sqrt{2\cosh{\pi{M}}}}
\left(
{e^{\frac{\pi{M}}{2}}}\Psi^{R}_{2,M,k_y,k_z}
-\epsilon{e^{-\frac{\pi{M}}{2}}}\Psi^{R,c}_{1,M,k_y,-k_z}
\right) \ ,
\end{align}
where $M>0$. The orthonormality relation is similar
to~(\ref{eq:psi-ON}). We have introduced the parameter $\epsilon$ in
(\ref{eqn:m-wmode12}) to label the spin structure: $\epsilon=1$
(respectively $-1$) gives spinors that are periodic (antiperiodic) in
the globally-defined vierbein~(\ref{eqn:rotet}).

The expansion of the field in the $W$-modes reads 
\begin{gather}
\Psi
=\sum_{n}\int_{0}^{\infty}dM\int_{-\infty}^{\infty}dk_y 
\left(
c_{1,M,k_y,k_z}W^R_{1,M,k_y,k_z}
+c_{2,M,k_y,k_z}W^R_{2,M,k_y,k_z}
\right.
\nonumber 
\\
\label{eqn:Wexp2}                                                   
\left.
\mbox{}+d^\dagger_{1,M,k_y,k_z}W^{R,c}_{1,M,k_y,k_z}
+d^\dagger_{2,M,k_y,k_z}W^{R,c}_{2,M,k_y,k_z}
\right) \ ,
\end{gather}
The annihilation and creation operators in (\ref{eqn:Wexp2}) satisfy
the usual anticommutation relations, and $|0_-\rangle$ is the state
annihilated by the annihilation operators. It follows that the
Bogolubov transformation between the Rindler-modes and the $W$-modes
reads
\begin{align}
a_{1,M,k_y,k_z}= 
&\ 
\frac{1}{\sqrt{2\cosh(\pi{M})}}
\left(
e^{\frac{\pi{M}}{2}}c_{1,M,k_y,k_z}
-\epsilon{e^{-\frac{\pi{M}}{2}}}d^{\dagger}_{2,M,k_y,-k_z}
\right) \ ,
\nonumber
\\
a_{2,M,k_y,k_z}= 
&\ 
\frac{1}{\sqrt{2\cosh(\pi{M})}}
\left(
e^{\frac{\pi{M}}{2}}c_{2,M,k_y,k_z}
+\epsilon{e^{-\frac{\pi{M}}{2}}}d^{\dagger}_{1,M,k_y,-k_z}
\right) \ ,
\nonumber
\\
b^\dagger_{1,M,k_y,k_z}= 
&\ 
\frac{1}{\sqrt{2\cosh(\pi{M})}}
\left(
e^{\frac{\pi{M}}{2}}d^\dagger_{1,M,k_y,k_z}
-\epsilon{e^{-\frac{\pi{M}}{2}}}c_{2,M,k_y,-k_z}
\right) \ , 
\nonumber
\\
\label{eqn:bog3}
b^\dagger_{2,M,k_y,k_z}= 
&\ 
\frac{1}{\sqrt{2\cosh(\pi{M})}}
\left(
e^{\frac{\pi{M}}{2}}d^\dagger_{2,M,k_y,k_z}
+\epsilon{e^{-\frac{\pi{M}}{2}}}c_{1,M,k_y,-k_z}
\right) \ , 
\end{align}
and the expansion of $|0_-\rangle$ in terms of the Rindler-excitations
is
\begin{equation}
\label{eqn:geonitmink}
|0_-\rangle
=\prod_{M,k_y,k_z}\frac{1}{({e^{-2\pi{M}}+1)^{\frac{1}{2}}}}
\sum_{q=0,1}(-\epsilon)^qe^{-\pi{M}q}
|q\rangle_{1,M,k_y,k_z}
|q\rangle_{2,M,k_y,-k_z} \ ,
\end{equation}
where 
\begin{align}
|q\rangle_{1,M,k_y,k_z}= &\ a^{\dagger}_{1,M,k_y,k_z}|0_{R_-}\rangle \ ,
\nonumber
\\
|q\rangle_{2,M,k_y,k_z}= &\ b^{\dagger}_{2,M,k_y,k_z}|0_{R_-}\rangle \ ,
\end{align}

From (\ref{eqn:geonitmink}) it is seen that $|0_-\rangle$ is an
entangled state of Rindler-excitations, and the correlations are
between a particle and an antiparticle with opposite eigenvalues
of~$k_z$. As all the excitations are in the unique Rindler
wedge~$R_-$, the expectation values of generic operators in $R_-$ are
not thermal. However, for any operator that only couples to one member
of each correlated pair in~$R_-$, the expectation values are
indistinguishable from those in the corresponding state $|0_0\rangle$
in~$R_0$, indicating thermality in the standard Unruh
temperature. This is the case for example for any operator that only
couples to excitations with a definite sign of~$k_z$. In particular,
the number operator expectation values are indistinguishable
from~(\ref{eqn:number}), and it can be argued from the isometries as
in
\cite{lm:geon} 
that the experiences of a Rindler-observer become asymptotically
thermal at early or late times.

A key result of our analysis is that while both spin structures on
$M_-$ induce a state in the same spin structure in~$R_-$, the explicit
appearance of $\epsilon$ in (\ref{eqn:geonitmink}) shows that the two
states differ. A~Rindler-observer in $R_-$ can therefore in principle
detect the spin structure on $M_-$ from the nonthermal
correlations. How these correlations could be detected in practice,
for example by particle detectors with a local coupling to the field,
presents an interesting question for future work. As the restriction
of $|0_-\rangle$ to $R_-$ is not invariant under the Killing
vector~$\partial_{\eta}$, and as the isometry arguments show that the
correlations disappear in the limit of large~$|\eta|$, investigating
this question would require a particle detector formalism that can
accommodate time dependent situations~\cite{sc:causaldet}.

\section{Stress-energy on $M_0$ and $M_-$ for massless spinors} 
\label{sec-Exp}

In this section we find the stress-energy expectation value for
massless spinors in the Minkowski-like vacua on $M_0$ and~$M_-$. 

Starting with the spinors of section \ref{sec-quant}, we set
$m=0$ and adopt the Weyl (chiral) representation of the
$\gamma$-matrices~\cite{it-zub}. 
Writing the 4-component spinor as 
$\psi=\bigl(
\begin{smallmatrix}
\psi_R \\ \psi_L
\end{smallmatrix}\bigr)$, 
the left-handed and right-handed 2-component spinors $\psi_L$ and
$\psi_R$ then decouple, and it suffices to consider the stress-energy
individually for each. This question was addressed
by Banach and Dowker \cite{bd:em} in a class of spatially compact 
flat spacetimes that
includes quotients of 
$M_0$ and~$M_-$. As the quotients are handled by taking image
sums, the stress-energy on $M_0$ and $M_-$ is 
obtained from the results in \cite{bd:em} by dropping the summations
that arise from the further quotients and matching the notation of
\cite{bd:em} to ours.\footnote{There is a
discrepancy in
\cite{bd:em} in the definition of the matrix $S(x)$ between the
Appendix and the bulk of the paper, one being the inverse of the
other, and this affects the sign in the last factor in equation (60)
therein and the related discussion at the top of page 2558. We agree
however with the final stress-energy results in~\cite{bd:em}.} The
remaining sums are over a single integer: The sums in the diagonal
components are purely numerical with well-known
values~\cite{zw:zwillinger}, and the sums in the nondiagonal
components on $M_-$ can be evaluated by residues (see e.g.\
\cite{spiegel-complex}, Chapter~7). We summarise the results in
Table~\ref{table:T-components}.

\begin{table}[t!]
\begin{center}
\begin{tabular}{|c|c|c|c|}
\hline
& 
$M_0$, $\epsilon=1$ 
& 
$M_0$, $\epsilon=-1$ 
& 
$M_-$ 
\\
\hline
$\langle{T_{00}}\rangle$ 
& 
$\displaystyle \frac{\pi^2}{720 a^4}$ 
&
$\displaystyle -\frac{7 \pi^2}{5760 a^4}$ 
&
$\displaystyle {-\frac{7 \pi^2}{5760 a^4}}^{\vphantom{A^A}}_{\vphantom{A_A}}$ 
\\
$\langle{T_{rr}}\rangle=\langle{T_{\hat{\phi}\hat{\phi}}}\rangle$ 
& 
$\displaystyle - \frac{\pi^2}{720 a^4}$ 
& 
$\displaystyle \frac{7 \pi^2}{5760 a^4}$ 
&
$\displaystyle {\frac{7 \pi^2}{5760 a^4}}^{\vphantom{A^A}}_{\vphantom{A_A}}$ 
\\
$\langle{T_{zz}}\rangle$ 
& 
$\displaystyle \frac{\pi^2}{240 a^4}$ 
&
$\displaystyle -\frac{7 \pi^2}{1920 a^4}$ 
&
$\displaystyle {-\frac{7 \pi^2}{1920
a^4}}^{\vphantom{A^A}}_{\vphantom{A_A}}$ 
\\
$\langle{T_{\hat{\phi}z}}\rangle$ 
& 
0 
& 
0 
&
$\displaystyle {-\frac{\epsilon\pi^2}{64a^4}\frac{d}{dq}
\!
\left(\frac{\sinh{q}}{q\cosh^2 \! {q}}\right)
}^{\vphantom{A^A}}_{\vphantom{A_{\displaystyle A}}}$
\\
\hline
\end{tabular}
\end{center}
\caption{The nonvanishing components of 
$\langle{T_{\mu\nu}}\rangle$ for a left-handed two-component spinor in
the Minkowski-like vacua on $M_0$ and $M_-$ in the orthonormal frame
$\{dt,dr,\omega^{\hat{\phi}},dz\}$, where $x=r\cos{\phi}$,
$y=r\sin{\phi}$ and $\omega^{\hat{\phi}} = r d\phi$, and $q=\pi
(r/a)$. On $M_0$, $\epsilon=1$ (respectively $-1$) indicates spinors
that are periodic (antiperiodic) in the nonrotating
vierbein~(\ref{eqn:minkvier}). On~$M_-$, $\epsilon=1$ (respectively
$-1$) indicates spinors that are periodic (antiperiodic) in the
vierbein~(\ref{eqn:rotet}). The values for a right-handed spinor are
identical. Note that the stress-energy is traceless in all cases.}
\label{table:T-components}
\end{table}

The results in Table \ref{table:T-components} show that the
stress-energy does not distinguish right-handed and left-handed
spinors, but it distinguishes $M_0$ from $M_-$ and the two spin
structures on each. On $M_0$ the $\epsilon =-1$ spin structure is
energetically preferred, while on $M_-$ both spin structures have the
same energy density. On $M_0$ the stress-energy tensor is diagonal and
invariant under all isometries. On $M_-$ the stress-energy is
invariant under all the continuous isometries, as it by construction
must be, but there is now a nonzero shear component
$\langle{T_{\hat{\phi}z}}\rangle$ whose sign is that of~$\epsilon$,
and this sign changes under isometries that reverse the spatial
orientation. 
On~$M_-$, the spatial 
orientation determined by the spin structure can thus be detected from
the shear part of the stress-energy. Note that the 
shear part vanishes at $r=0$ and tends to zero exponentially as $r
\to \infty$, but numerical evidence shows that there is a range in $r$
where this shear part is in fact the dominant part of the
stress-energy.

\section{Massive spinors on the ${\mathbb{RP}}^3$ geon}
\label{sec-hawking}

In this section we analyse the Hawking effect on the ${\mathbb{RP}}^3$
geon for the massive Dirac field. Subsection~\ref{sec-krusk} recalls
some properties of the geon geometry and fixes the notation, and the
Boulware vacuum in the exterior region is presented in
subsection~\ref{sec-boulware}. The main results are in
section~\ref{sec-hhvac}, where the Hartle-Hawking-like vacuum for each
spin structure is constructed and analysed.

\subsection{Kruskal spacetime and the ${\mathbb{RP}}^3$ geon}
\label{sec-krusk}

In the notation of~\cite{lm:geon}, the Kruskal 
metric in the Kruskal coordinates 
$(T,X,\theta,\phi)$ reads 
\begin{equation}
{ds}^2=
\frac{32M^3}{r}e^{-r/(2M)}
({dT}^2-{dX}^2)
-r^2({d\theta}^2+{\sin^2{\theta}}{d\phi}^2) \ , 
\end{equation}
where $M>0$, $T^2 - X^2 < 1$ and $r$ is determined as a function of
$T$ and $X$ by $T^2 - X^2 = 1 - r/(2M)$. The manifold consists of the
right and left exteriors, denoted respectively by $R$ and $L$, and the
future and past interiors, denoted respectively by $F$ and $P$,
separated from each other by the bifurcate Killing horizon at $|T| =
|X|$. The four regions are individually covered by the Schwarzschild
coordinates $(t,r,\theta,\phi)$, in which the metric reads
\begin{equation}
{ds}^2=
\left(1-\frac{2M}{r}\right){dt}^2
-\frac{{dr}^2}{(1-\frac{2M}{r})}
-r^2({d\theta}^2+{\sin^2{\theta}}{d\phi}^2) \ , 
\label{eq:schw}
\end{equation}
where $2M < r < \infty$ in the exteriors and $0 < r < 2M$ in the
interiors. The coordinate transformation in $R$ is
\begin{align}
T= &\ \left(\frac{r}{2M}-1\right)^{\frac{1}{2}}e^{r/(4M)}
\sinh{\left(\frac{t}{4M}\right)} \ ,
\nonumber
\\
X= &\ \left(\frac{r}{2M}-1\right)^{\frac{1}{2}}e^{r/(4M)}
\cosh{\left(\frac{t}{4M}\right)} \ , 
\label{eq:schw-kruskal-transf}
\end{align}
and the transformations in the other regions are given in~\cite{MTW}. 
The exteriors are static, with the timelike Killing vector~$\partial_t$. 

The ${\mathbb{RP}}^3$ geon is the quotient of the Kruskal manifold 
under the $\mathbb{Z}_2$ isometry group generated by the map 
\begin{equation}
\label{eqn:involJ}
J:(T,X,\theta,\phi)\mapsto(T,-X,\pi-\theta,\phi+\pi) \ . 
\end{equation}
The construction is analogous to that of $M_-$ from $M_0$ in
subsection~\ref{sec-spacet}. Further discussion, including conformal
diagrams, can be found in~\cite{lm:geon}.

As the Kruskal manifold has spatial topology
${\mathbb{R}}\times{S^2}$, it is simply connected and has a unique
spin structure. The quotient construction implies \cite{hatcher} that
the geon has fundamental group~${\mathbb{Z}}_2$ and admits two spin
structures. As in section~\ref{sec-unruh}, we describe these spin
structures in terms of periodic and antiperiodic boundary conditions
for spinors in a specified vierbein. On Kruskal, a standard reference
vierbein is
\begin{eqnarray}
\label{eqn:kruskvier} 
V_0=\partial_T & V_1=\partial_X 
\nonumber 
\\ 
V_2=\partial_{\theta} & V_3=\partial_{\phi} \ .  
\end{eqnarray}
A second useful vierbein is 
\begin{eqnarray} 
V_0 & = & \partial_T 
\nonumber 
\\
V_1 & = & \cos{\phi}\partial_X+\sin{\phi}\partial_{\theta} 
\nonumber 
\\
V_2 & = & -\sin{\phi}\partial_X+\cos{\phi}\partial_{\theta} 
\nonumber 
\\
V_3 & = & \partial_{\phi} \ , 
\label{eq:geon-bein} 
\end{eqnarray}
which rotates by $\pi$ in the $(X,\theta)$ tangent plane as $\phi$
increases by $\pi$ and is invariant under~$J$. The vierbein
(\ref{eq:geon-bein}) is well defined on the geon, and when the spinors
are written in it, the two spin structures correspond to respectively
periodic and antiperiodic boundary conditions as
$\phi\mapsto{\phi+\pi}$. Both (\ref{eqn:kruskvier}) and
(\ref{eq:geon-bein}) are singular at $\theta=0$ and $\theta=\pi$, but
these coordinate singularities on the sphere 
can be handled by usual methods
and will not affect our discussion. 

In practice, we will work in the standard
vierbein~(\ref{eqn:kruskvier}). The boundary conditions appropriate
for the two geon spin structures will be found by the method-of-images
technique of the Appendix of~\cite{bd:em}.

\subsection{The Boulware vacuum}
\label{sec-boulware}

In this subsection we review the construction of the Boulware vacuum
in one exterior region~\cite{bo:boulware}. While this vacuum as such
is well known, we will need to decompose the field in a novel basis in
order to make contact with the geon in subsection~\ref{sec-hhvac}.

We work in the Schwarzschild coordinates (\ref{eq:schw}), with $r>2M$, 
and in the adapted vierbein, 
\begin{equation}
\label{eqn:schwarzvier}
V^{\mu}_{a}
=\mathrm{diag}
\left(
\frac{1}{\left(1-\frac{2M}{r}\right)^{\frac{1}{2}}},
\left(1-\frac{2M}{r}\right)^{\frac{1}{2}},
\frac{1}{r},
\frac{1}{r\sin{\theta}}
\right) \ . 
\end{equation}
The Diraq equation (\ref{eqn:gendir}) becomes
\begin{eqnarray}
\left[
m
+\frac{\gamma^2}
{ir\sin^{\frac{1}{2}}\theta}
\partial_{\theta}\sin^{\frac{1}{2}}\theta 
+\frac{\gamma^3}
{ir\sin{\theta}}
\partial_{\phi}
+\frac{\gamma^0}
{i(1-\frac{2M}{r})^{\frac{1}{2}}}
\partial_t 
\right.
\nonumber 
\\
\label{eqn:dirinschwarz}
\left.
\mbox{}
+\frac{\gamma^1}
{ir}(1-\frac{2M}{r})^{\frac{1}{4}}
\partial_r(1-\frac{2M}{r})^{\frac{1}{4}}r
\right]
\psi=0 \ ,
\end{eqnarray}
where the $\gamma$
matrices are flat space $\gamma$'s. We adopt the representation 
$$\begin{array}{cc}\gamma^0=\left(\begin{array}{cccc}
0 & 0 & 1 & 0 \\
0 & 0 & 0 & 1 \\
1 & 0 & 0 & 0 \\
0 & 1 & 0 & 0 \\
\end{array} \right) \ , &
\gamma^1=\left(\begin{array}{cccc}
0 & 0 & -1 & 0 \\
0 & 0 & 0 & -1 \\
1 & 0 & 0 & 0 \\
0 & 1 & 0 & 0 \\
\end{array}\right) \ , \end{array} $$
\begin{equation}\label{eqn:gammarep}
\begin{array}{cc}\gamma^2=\left(\begin{array}{cccc}
-i & 0 & 0 & 0 \\
0 & i & 0 & 0 \\
0 & 0 & i & 0 \\
0 & 0 & 0 & -i \\
\end{array}\right) \ , &
\gamma^3=\left(\begin{array}{cccc}
0 & -i & 0 & 0 \\
-i& 0 & 0 & 0 \\
0 & 0 & 0 & i \\
0 & 0& i & 0 \\
\end{array}\right) \ , 
\end{array} 
\end{equation}
which has the advantage that charge conjucation takes the simple form 
\begin{equation}
\psi^c=
\left(
\begin{array}{cccc}
1 & 0 & 0 & 0 \\
0 & 1 & 0 & 0 \\
0 & 0 & -1 & 0 \\
0 & 0 & 0 & -1 \\
\end{array}
\right)
\psi^* \ ,
\label{eq:chargeconjS}
\end{equation}
where ${}^*$ again stands for complex conjugation. 

We use the separation ansatz 
\begin{equation}
\label{eqn:dirsolution}
\psi_{\omega,k',m'}(t,r,\theta,\phi)
=N\frac{e^{-i\omega{t}}}{r(1-\frac{2M}{r})^{\frac{1}{4}}}
\left(
\begin{array}{c}
F(r)Y^{k'}_{m'}(\theta,\phi) \\
G(r)Y^{k'}_{m'}(\theta,\phi) \\
\end{array}
\right)_{\omega{k'}} \ ,
\end{equation}
where $\omega>0$ for modes that are positive frequency with respect to
the Killing vector $\partial_t$ and the spinor spherical harmonics
$Y^{k'}_{m'}(\theta,\phi)$ are as constructed
in~\cite{bo:boulware}. The radial functions then satisfy
\begin{subequations}
\label{eqn:rad12}
\begin{align}
\label{eqn:rad1}
\left(
1-\frac{2M}{r}
\right)
\partial_rF-i\omega{F}= 
&\ 
\left(
1-\frac{2M}{r}
\right)
^{\frac{1}{2}}
\left(
\frac{k'}{r}-im
\right)
G \ , 
\\
\label{eqn:rad2}
\left(
1-\frac{2M}{r}
\right)
\partial_rG+i\omega{G}= 
&\ 
\left(
1-\frac{2M}{r}
\right)
^{\frac{1}{2}}
\left(
\frac{k'}{r}+im
\right)
F \ . 
\end{align}
\end{subequations}

Following Chandrasekhar~\cite{ch:chbook}, we reduce the radial
equations (\ref{eqn:rad12}) to a pair of Schr\"odinger-like
equations. Writing
\begin{equation}
\left(
\begin{array}{c} 
F(r) 
\\
G(r) 
\end{array}
\right)
=\left(
\begin{array}{c} 
\frac{1}{2}(Z_++Z_-)e^{(-\frac{i}{2}\tan^{-1}(\frac{mr}{k'}))} 
\\ 
\frac{1}{2}(Z_+-Z_-)e^{(\frac{i}{2}\tan^{-1}(\frac{mr}{k'}))}
\end{array}
\right) \ ,
\label{eq:FGvZ}
\end{equation}
we find that  $Z_{\pm}$ satisfy 
\begin{equation}
\label{eqn:zplusminus} 
\left(
\frac{d}{d\hat{r}_*}{\mp}W\right)Z_{\pm}
=i\omega{Z_{\mp}} \ , 
\end{equation}
where $\hat{r}^*=r^*+\frac{1}{2\omega}\tan^{-1}(\frac{mr}{k'})$, 
$r^*=r+2M\ln(|r-2M|/2M)$ and
\begin{equation}
W=
\frac{(r^2-2Mr)^{\frac{1}{2}}(k'^2+m^2r^2)^{\frac{3}{2}}}{r^2(k'^2+m^2r^2)
+\frac{k'm}{2\omega}(r^2-2Mr)} \ .
\end{equation}
$Z_{\pm}$ hence satisfy the one-dimensional wave equations
\begin{equation}
\label{eqn:scat}
\left(
\frac{d^2}{d\hat{r}^2_*}
+\omega^2\right)Z_{\pm}=V_{\pm}Z_{\pm} \ ,
\end{equation}
where 
\begin{equation}
V_{\pm}=W^2\pm\frac{dW}{d\hat{r}_*} 
\ . 
\end{equation}

Suppose first $\omega^2>m^2$ in~(\ref{eqn:scat}), in which case there
are two linearly independent delta-normalisable solutions for
each~$\omega$. One way to break this degeneracy would be to choose
solutions that have the scattering theory asymptotic form,
\begin{subequations}
\label{eqn:zleftright}
\begin{equation}
\label{eqn:zleft}
\stackrel{\leftarrow}{Z_\pm}
=
\left\{
\begin{array}{cc}
\stackrel{\leftarrow}{B_\pm}e^{-i\omega{\hat{r_*}}} 
&  
\hat{r_*}\rightarrow{-\infty} 
\\ 
e^{-i(p\hat{r_*}+\frac{Mm^2}{p}\ln(\frac{\hat{r_*}}{2M}))}
+\stackrel{\leftarrow}{A_\pm}e^{i(p\hat{r_*}
+\frac{Mm^2}{p}\ln(\frac{\hat{r_*}}{2M}))} 
& 
\hat{r_*}\rightarrow{\infty} \ \ ,
\end{array}
\right. 
\end{equation}
\begin{equation}
\label{eqn:zright}
\stackrel{\rightarrow}{Z_\pm}
=
\left\{
\begin{array}{cc}e^{i\omega\hat{r_*}}
+\stackrel{\rightarrow}{A_\pm}e^{-i\omega\hat{r_*}} 
&  
\hat{r_*}\rightarrow{-\infty} 
\\ 
\stackrel{\rightarrow}{B_\pm}e^{i(p\hat{r_*}
+\frac{Mm^2}{p}\ln(\frac{\hat{r_*}}{2M}))} 
& 
\hat{r_*}\rightarrow{\infty} \ \ , 
\end{array}
\right. 
\end{equation}
\end{subequations}
where $p=\sqrt{(\omega^2-m^2)}$. $\stackrel{\leftarrow}{Z_\pm}$ is
purely ingoing at the horizon and $\stackrel{\rightarrow}{Z_\pm}$ is
purely outgoing at infinity, and usual scattering theory Wronskians
yield relations between the transmission and reflection
coefficients. From (\ref{eqn:zplusminus}) it further follows that
$\stackrel{\leftarrow}{B_+}=-\stackrel{\leftarrow}{B_-}$ and
$\stackrel{\rightarrow}{A_+}=-\stackrel{\rightarrow}{A_-}$, which will
fix the form of a radial mode (\ref{eq:FGvZ}) that is purely ingoing
at the horizon and a radial mode that is purely outgoing at
infinity. However, to be able to handle the geon in
subsection~\ref{sec-hhvac}, we will need modes that transform simply
under charge conjugation~(\ref{eq:chargeconjS}) and under $J$ \ref{eqn:involJ} when 
continued analytically into the $F$ region. Using the Wronskian
properties of the reflection and transmission coefficients in
(\ref{eqn:zleftright}) and the properties of the spinor spherical
harmonics~\cite{bo:boulware}, 
we find after considerable effort that a
convenient set of positive frequency Boulware modes is
$\{\Psi^{\pm}_{\omega,k',m'}\}$, given by
\begin{subequations}
\label{eqn:posmode12}
\begin{align}
\label{eqn:posmode1}
\Psi^+_{\omega,m',k'}
&=\frac{
e^{-\frac{i\pi}{2}(j+m'+(\frac{k'}{|k'|}-1)/2)}
e^{-i\omega{t}}
}
{r(1-\frac{2M}{r})^{\frac{1}{4}}}
{{
\left(
\begin{array}{c}
u(r)Y^{k'}_{m'}(\theta,\phi) 
\\
v(r)Y^{k'}_{m'}(\theta,\phi) 
\\
\end{array}
\right)
}}^+_{\omega{k'}} \ ,
\\
\label{eqn:posmode2}
\Psi^-_{\omega,m',k'}
&=\frac{
e^{-\frac{i\pi}{2}(j-m'+(\frac{k'}{|k'|}-1)/2)}
e^{-i\omega{t}}
}
{r(1-\frac{2M}{r})^{\frac{1}{4}}}
{{
\left(
\begin{array}{c}
u(r)Y^{k'}_{m'}(\theta,\phi) 
\\
v(r)Y^{k'}_{m'}(\theta,\phi) 
\\
\end{array}
\right)
}}^-_{\omega{k'}} \ ,
\end{align}
\end{subequations}
where $\omega>m$, the radial functions with superscript ${}^+$ are
specified by the horizon asymptotic behaviour
\begin{subequations}
\label{eq:deg-nearhor}
\begin{align}
\left(
\begin{array}{c}u(r) 
\\
v(r) 
\\
\end{array}
\right)^+_{\omega{k'}}
&=
\frac{1}{\sqrt{4\pi}}
\left\{
\sqrt{1+\sqrt{1-|\stackrel{\rightarrow}{A}|^2}}
\left(
\begin{array}{c}
1 
\\
0 
\\
\end{array}
\right)
e^{i\omega{r_*}}
\vphantom{
\frac{\stackrel{\rightarrow}{A_+}}
{\sqrt{1+\sqrt{1-|\stackrel{\rightarrow}{A}|^2}}}
}
\right.
\nonumber
\\
\label{eqn:+mode}
&
\hspace{5ex}
\left.
+\frac{\stackrel{\rightarrow}{A_+}}
{\sqrt{1+\sqrt{1-|\stackrel{\rightarrow}{A}|^2}}}
\left(
\begin{array}{c}
0 
\\
1 
\\
\end{array}
\right)
e^{-i\omega{r_*}}
\right\},
\ \ \ \ \ \hat{r}^*\to{-\infty} 
\ , 
\end{align}
and
\begin{equation}
\label{eqn:-mode}
\left(
\begin{array}{c}
u(r) 
\\
v(r) 
\\
\end{array}
\right)^-_{\omega{k'}}
=
\left(
\begin{array}{cc} 
0 & 1
\\ 
1 & 0 
\end{array}
\right)
\left(
\begin{array}{c}
u(r) 
\\
v(r) 
\\
\end{array}
\right)^{+*}_{\omega{k'}} \ . 
\end{equation}
\end{subequations}
The key property for charge conjugation is~(\ref{eqn:-mode}). 
The modes are complete for $\omega^2 > m^2$ 
and delta-orthonormal in the Dirac inner product 
\begin{equation}
\label{eqn:innerprod}
\langle\psi_1,\psi_2\rangle
=\int_{\mathrm{angles}}\sin\theta\,
{d\theta}\,
{d\phi}\,
\int^{\infty}_{2M}\frac{r^2}{\left(1-\frac{2M}{r}\right)^{\frac{1}{2}}}
\psi_1^{\dagger}\psi_2\,dr \ ,
\end{equation}
taken on a constant $t$ hypersurface. 

Suppose then $0<\omega^2<m^2$ in~(\ref{eqn:scat}). There is now only
one linearly independent delta-normalisable solution for
each~$\omega$. This solution vanishes at infinity and has at the
horizon the behaviour
\begin{equation}
Z_{\pm}=a_{\pm}\cos(\omega\hat{r}^*+\delta_{\pm})  \ ,
\ \ \ 
\hat{r}^*\to{-\infty} 
\ , 
\end{equation}
where $a_{\pm}$, and $\delta_{\pm}$ are real constants. Physically
these solutions correspond to particles that do not reach
infinity. Proceeding as above, we find that a convenient set of
positive frequency Boulware modes, complete for $0<\omega^2<m^2$ and
delta-orthonormal in the Dirac inner product~(\ref{eqn:innerprod}), is
\begin{equation}
\label{eqn:posmode3}
\psi_{\omega,k',m'}(t,r,\theta,\phi)
=e^{-\frac{i\pi}{2}(j+|m'|+(1-\frac{k'}{|k'|})/2)}
\frac{e^{-i\omega{t}}}{r(1-\frac{2M}{r})^{\frac{1}{4}}}
\left(
\begin{array}{c}
F(r)Y^{k'}_{m'}(\theta,\phi) \\
G(r)Y^{k'}_{m'}(\theta,\phi) \\
\end{array}
\right)_{\omega{k'}} \ , 
\end{equation}
where $0<\omega < m$ and the radial functions are specified by the
horizon asymptotic behaviour
\begin{equation}
\label{eqn:wmodewm}
\left(
\begin{array}{c}
F(r) 
\\
G(r) 
\\
\end{array}
\right)_{\omega{k'}}
=
\frac{1}{\sqrt{2\pi}}
\left\{
\left(
\begin{array}{c} 
e^{i\delta_+} \\
0 \\
\end{array}
\right)
e^{i\omega{r_*}}+
\left(
\begin{array}{c}
0 \\
e^{-i\delta_+} \\
\end{array}
\right)
e^{-i\omega{r_*}}
\right\} \ ,
\ \ \ 
\hat{r}^*\mapsto{-\infty} \ . 
\end{equation}

Up to this point we have used the Schwarzschild
vierbein~(\ref{eqn:schwarzvier}). To make contact with the geon in
subsection~\ref{sec-hhvac}, we need to express the modes in a vierbein
that is regular at the horizons. We therefore now transform our modes
to the Kruskal
vierbein (\ref{eqn:kruskvier}) by the spinor transformation
$\psi\mapsto{e^{\frac{t}{8M}\gamma^0\gamma^1}\psi}$. We suppress the
explicit transformed expressions and 
continue to use the same symbols for the mode
functions.

We are now ready to quantise. The field is expanded in our orthonormal
modes and their charge conjugates as
\begin{eqnarray}       
\Psi                   
& = & 
\sum_{m',k'}
\int_{0}^{m}d\omega
\left(
a_{\omega,m',k'}\psi_{\omega,m',k'}
+b^{\dagger}_{\omega,m',k'}\psi^c_{\omega,m',k'} 
\right) 
\nonumber 
\\ 
&   & 
\mbox{}
+\sum_{m',k'}
\int_{m}^{\infty}d\omega
\left(
a_{+,\omega,m',k'}\Psi^+_{\omega,m',k'}+
a_{-,\omega,m',k'}\Psi^-_{\omega,m',k'}
\right. 
\nonumber 
\\
\label{eqn:boulexp}
&   & 
\hspace{17ex}
\left.
\mbox{}
+b^\dagger_{+,\omega,m',k'}\Psi^{+,c}_{\omega,m',k'}
+b^\dagger_{-,\omega,m',k'}\Psi^{-,c}_{\omega,m',k'}
\right) \ ,  
\end{eqnarray}
where the annihilation and creation operators satisfy the usual
anticommutation relations. The vacuum annihilated by the annihilation
operators is the Boulware vacuum~$|0_B\rangle$. $|0_B\rangle$ is by
construction the state void of particles with respect to the
Schwarzschild Killing time.

\subsection{The Hartle-Hawking like vacuum and 
Bogolubov transformation on the geon}
\label{sec-hhvac}

In this subsection we decompose the geon Hartle-Hawking-like vacuum
into Boulware excitations. We use the analytic continuation method,
following closely subsection~\ref{sec-bog2}. 

The Hartle-Hawking vacuum on Kruskal is defined by mode functions that
are purely positive frequency with respect to the horizon generators
and hence analytic in the lower half-plane in the complexified Kruskal
time~$T$. It follows that on Kruskal we can construct $W$-modes whose
vacuum is the Hartle-Hawking vacuum by analytically continuing the
Boulware-modes across the horizons in the lower half-plane in~$T$. The
quotient from Kruskal to the geon defines in each spin structure on
the geon the Hartle-Hawking-like vacuum $|0_G\rangle$, by restriction
to the Kruskal $W$-modes that are invariant under the
map $J$~(\ref{eqn:involJ}). Our task is to find these modes. 

Let us denote by $W^F$ the restriction of the sought-for $W$-modes to
the region~$F$ and by $\psi^F$ the analytic continuation of Boulware
modes from $R$ to~$F$, expressed in the rotating Kruskal
vierbein~(\ref{eq:geon-bein}). As this vierbein is invariant
under~$J$, we can set 
\begin{equation}
\label{eq:rot-Wgeon-trans}
W^F(T,X,\theta,\phi)
=\psi^F(T,X,\theta,\phi)
+\epsilon\psi^F(T,-X,\pi-\theta,\phi+\pi) \ ,
\end{equation}
where $\epsilon=1$ (respectively $-1$) for spinors that are periodic
(antiperiodic) in this vierbein. As the transformation from this
vierbein to the standard Kruskal vierbein (\ref{eqn:kruskvier}) is a
rotation by $-\pi$ in the $(X,\theta)$ tangent plane as $\phi$
increases by~$\pi$, the corresponding spinor transformation is 
\begin{equation} 
W^F(T,X,\theta,\phi)
\mapsto
W^F_s(T,X,\theta,\phi)
:= {e^{\frac{\phi\gamma^1\gamma^2}{2}}}W^F(T,X,\theta,\phi)
\ , 
\end{equation}
where the subscript $s$ refers to the standard
vierbein~(\ref{eqn:kruskvier}). In the standard vierbein,
(\ref{eq:rot-Wgeon-trans}) hence becomes 
\begin{equation}
\label{eqn:methofim}
W^F_s(T,X,\theta,\phi)
=\psi^F_s(T,X,\theta,\phi)
+\epsilon{e^{-\frac{\pi\gamma^1\gamma^2}{2}}}
\psi^F_s(T,-X,\pi-\theta,\phi+\pi) \ .
\end{equation}
We can therefore employ the condition~(\ref{eqn:methofim}). It can be
verified that the functions have the correct transformation
properties also in the $P$ region of Kruskal.

The computations are lengthy but straightforward, using the explicit
coordinate transformation (\ref{eq:schw-kruskal-transf}) 
and the near-horizon behaviour in
(\ref{eq:deg-nearhor}) and~(\ref{eqn:wmodewm}). 
Continued back to~$R$, we find that the 
$W$-modes with $\omega>m$ are 
\begin{align}
W^+_{\omega,k',m'}(t,r,\theta,\phi)
&=\frac{1}{\sqrt{2\cosh(4\pi{M}\omega)}}
\left(
e^{2\pi{M}\omega}\Psi^+_{\omega,k',m'}
+\epsilon{e^{-2\pi{M}\omega}}\Psi^{-,c}_{\omega,k',-m'}
\right) \ ,
\nonumber
\\
W^-_{\omega,k',m'}(t,r,\theta,\phi)
&=\frac{1}{\sqrt{2\cosh(4\pi{M}\omega)}}
\left(
e^{2\pi{M}\omega}\Psi^-_{\omega,k',m'}
-\epsilon{e^{-2\pi{M}\omega}}\Psi^{+,c}_{\omega,k',-m'}
\right) \ ,
\end{align}
and those with $0<\omega<m$ are 
\begin{equation}
W_{\omega,k',m'}(t,r,\theta,\phi)
=\frac{1}{\sqrt{2\cosh(4\pi{M}\omega)}}
\left(
e^{2\pi{M}\omega}\psi_{\omega,k',m'}
+\epsilon\frac{m'}{|m'|}{e^{-2\pi{M}\omega}}\psi^{c}_{\omega,k',-m'}
\right) \ . 
\label{eq:W-geon-deg}
\end{equation}
These modes are appropriately delta-orthonormal in the Dirac inner
product (\ref{eqn:innerprod}) in $R$ and hence also orthonormal in
the Dirac inner product on the geon. It can be verified that the
factors $m'/|m'|$ appearing in (\ref{eq:W-geon-deg}) cannot be
absorbed into the phase factors of the Boulware modes. 

On the geon, the expansion of the field in the $W$-modes reads 
\begin{eqnarray}
\Psi 
& = & 
\sum_{m',k'}
\int_{0}^{m}d\omega
\left(
c_{\omega,m',k'}W_{\omega,m',k'}
+d^{\dagger}_{\omega,m',k'}W^c_{\omega,m',k'} 
\right)
\nonumber 
\\ 
&   &
\mbox{}
+\sum_{m',k'}
\int_{m}^{\infty}d\omega
\left(
c_{+,\omega,m',k'}W^+_{\omega,m',k'}
+c_{-,\omega,m',k'}W^-_{\omega,m',k'} 
\right. 
\nonumber
\\
\label{eqn:boulexp2}
&   &
\hspace{17ex}
\left.
\mbox{}
+d^\dagger_{+,\omega,m',k'}W^{+,c}_{\omega,m',k'}
+d^\dagger_{-,\omega,m',k'}W^{-,c}_{\omega,m',k'}
\right) \ ,
\end{eqnarray}
with the usual anticommutation relations, and $|0_G\rangle$ is the state
annihilated by all the annihilaton operators. Comparison of
(\ref{eqn:boulexp}) and (\ref{eqn:boulexp2}) gives the Bogolubov 
transformation 
\begin{align}
a_{\omega,k',m'}= 
&\ 
\frac{1}{\sqrt{2\cosh(4\pi{M}\omega)}}
\left(
e^{{2\pi{M}\omega}}c_{\omega,k',m'}
+\epsilon\frac{m'}{|m'|}{e^{-{2\pi{M}}\omega}}d^{\dagger}_{\omega,k',-m'}
\right) \ ,
\nonumber 
\\
b^{\dagger}_{\omega,k',m'}= 
&\ 
\frac{1}{\sqrt{2\cosh(4\pi{M}\omega)}}
\left(
e^{{2\pi{M}\omega}}d^{\dagger}_{\omega,k',m'}
+\epsilon\frac{m'}{|m'|}{e^{-{2\pi{M}}\omega}}c_{\omega,k',-m'}
\right) \ ,
\nonumber
\\
a_{+,\omega,k',m'}= 
&\ 
\frac{1}{\sqrt{2\cosh(4\pi{M}\omega)}}
\left(
e^{2\pi{M}\omega}c_{+,\omega,k',m'}
-\epsilon{e^{-2\pi{M}\omega}}d^{\dagger}_{-,\omega,k',-m'}
\right) \ , 
\nonumber
\\
a_{-,\omega,k',m'}= 
&\ 
\frac{1}{\sqrt{2\cosh(4\pi{M}\omega)}}
\left(
e^{2\pi{M}\omega}c_{-,\omega,k',m'}
+\epsilon{e^{-2\pi{M}\omega}}d^{\dagger}_{+,\omega,k',-m'}
\right) \ ,
\nonumber
\\
b^\dagger_{+,\omega,k',m'}= 
&\ 
\frac{1}{\sqrt{2\cosh(4\pi{M}\omega)}}
\left(
e^{2\pi{M}\omega}d^\dagger_{+,\omega,k',m'}
-\epsilon{e^{-2\pi{M}\omega}}c_{-,\omega,k',-m'}
\right) \ ,
\nonumber
\\
b^\dagger_{-,\omega,k',m'}= 
&\ 
\frac{1}{\sqrt{2\cosh(4\pi{M}\omega)}}
\left(
e^{2\pi{M}\omega}d^\dagger_{-,\omega,k',m'}
+\epsilon{e^{-2\pi{M}\omega}}c_{+,\omega,k',-m'}
\right) \ . 
\end{align}
It follows that the expansion of $|0_G\rangle$ in the
Boulware-excitations is 
\begin{align}
|0_G\rangle 
& = 
\prod_{\substack{0<\omega < m \\ m',k'}}
\frac{1}{({e^{-8\pi{M}\omega}+1)^{\frac{1}{2}}}}
\sum_{q=0,1}
{\left(
\frac{\epsilon{m'}}{|m'|} 
\right)}^q
e^{-4\pi{M}\omega{q}}
(a^{\dagger}_{\omega,k',m'}b^{\dagger}_{\omega,k',-m'})^q
|0_B\rangle 
\nonumber
\\
& 
\hspace{3ex}
\times 
\prod_{\substack{\omega > m \\ m',k'}}
\frac{1}{({e^{-8\pi{M}\omega}+1)^{\frac{1}{2}}}}\!\!
\sum_{q=0,1}(-\epsilon)^qe^{-4\pi{M}\omega{q}}
|q\rangle_{+,\omega,m',k'}
|q\rangle_{-,\omega,-m',k'} 
\ ,
\label{eqn:geonvac}
\end{align}
where 
\begin{align}
|q\rangle_{+,\omega,m',k'}= 
&\ 
a^{\dagger}_{+,\omega,m',k'}|0_B\rangle \ ,
\nonumber
\\
|q\rangle_{-,\omega,m',k'}= 
&\ 
b^{\dagger}_{-,\omega,m',k'}|0_B\rangle \ .
\end{align}

A comparison of (\ref{eqn:geonvac}) and (\ref{eqn:geonitmink}) shows
that $|0_G\rangle$ is closely similar to the state $|0_-\rangle$
on~$M_-$, and the discussion at the end of subsection \ref{sec-bog2}
adapts directly here. $|0_G\rangle$~does not appear thermal to generic
static observers in~$R$, but it appears thermal in the standard
Hawking temperature ${(8\pi M)}^{-1}$ near the inifinity when probed
by operators that only couple to one member of each correlated pair
in~(\ref{eqn:geonvac}), such as operators that only couple to a
definite sign of the angular momentum quantum number~$m'$. In
particular, number operator expectation values are thermal, and the
isometry arguments of \cite{lm:geon} show that the experiences of any
static observer become asymptotically thermal in the large $|t|$
limit. 

The explicit appearance of $\epsilon$ in (\ref{eqn:geonvac}) shows
that the nonthermal correlations in $|0_G\rangle$ reveal the geon spin
structure to an observer in~$R$. This is a phenomenon that could not
have been anticipated just from the geometry of~$R$, which in its own
right has only one spin structure.

\section{Discussion}
\label{sec-conclu}

This paper has discussed thermal effects for the free Dirac field on
the ${\mathbb{RP}}^3$ geon and on a topologically analogous flat
spacetime $M_-$ via a Bogolubov transformation analysis. Compared with
the scalar field~\cite{lm:geon}, the main new issue with fermions is
that the spacetimes admit two inequivalent spin structures, and there
are hence two inequivalent Hartle-Hawking like vacua on the geon and
two inequivalent Minkowski-like vacua on~$M_-$. We showed that an
observer in the exterior region of the geon can detect both the
nonthermality of the Hartle-Hawking like state and the spin structure
of this state by suitable intereference measurements, and similar
results hold for a Rindler observer on~$M_-$. When probed with
suitably restricted operators, such as operators at asymptotically
late Schwarzschild (respectively Rindler) times, these states
nevertheless appear thermal in the usual Hawking (Unruh) temperature,
for the same geometric reasons as in the scalar
case~\cite{lm:geon}. We further computed the stress-energy expectation
value on $M_-$ in the massless limit, showing that the two spin
structures are distinguished by the sign of a nonvanishing shear
component. As a by-product of the analysis, we presented the Bogolubov
transformation for the Unruh effect for the massive Dirac field in
$(3+1)$-dimensional Minkowski space, complementing and correcting the
previous literature. 

It would be interesting to explore how to observe the nonthermal
correlations in the Hartle-Hawking like and Minkowski-like states by
particle detectors that couple to the field in some local fashion. As
the states in question are are not invariant under the
locally-defined Killing vectors with respect to which the thermal
properties arise, 
the deviations from thermality would need to be analysed in
a setting that can handle time-dependent detector
response~\cite{sc:causaldet}.

As a late time observer in the geon exterior sees a thermal state in
the usual Hawking temperature, the classical laws of black hole
mechanics lead the observer to assign to the geon the same entropy as
to a conventional Schwarzschild hole with the same mass. It was found
in \cite{lm:geon} that an attempt to evaluate the geon entropy by
path-integral methods leads to half of the Bekenstein-Hawking entropy
of a Schwarzschild hole of the same mass, and it was suggested
that state-counting computations of the geon entropy could shed light
on this discrepancy. Our work says little of what the full framework
of such a computation would be, but our work would presumably provide
part of the fermionic machinery in the computation. In particular, the
issue of the spin structure would need to be faced seriously: 
Does an entropy computation by state-counting need to count the two
spin structures as independent degrees of freedom?

\subsection*{Acknowledgements}

I would like to thank Jorma Louko for many useful discussions and
for reading the manuscript. Thanks also to Edward Armour and John
Barrett for useful comments and to Daniele Oriti for correspodence on
reference~\cite{or:oriti}. Comments from an anonymous referee helped
improve the presentation. This work was supported by the University of
Nottingham.

\end{document}